%% file: main_arxiv.tex
\documentclass[amssymb,prb,twocolumn,superscriptaddress,floats,showkeys,amsmath]{revtex4-2}

\usepackage{graphicx}
\usepackage{epstopdf}
\usepackage{hyperref}
\hypersetup{colorlinks,citecolor=blue, filecolor=blue ,linkcolor=blue , urlcolor=blue}

\def \FUW{University of Warsaw, Faculty of Physics, 02-093 Warsaw, Poland}
\def \Beihang{National Key Lab of Spintronics, International Innovation Institute, Beihang University,
Hangzhou 311115, China} 
\def \Cem{Department of Physics-NANOlight Center of Excellence, University of Antwerp, Antwerp B-2020, Belgium}
\def\Pakistan{Nanoscale Synthesis \& Research Laboratory, Department of Applied Physics, University of Karachi, Karachi 75270, Pakistan }

\begin{document}

\title{Raman scattering fingerprints of the charge density wave state \\ in one-dimensional NbTe$_4$}

\author{Natalia Zawadzka}
\email{Natalia.Zawadzka@fuw.edu.pl}
\affiliation{\FUW}
\author{Cem Sevik}
\affiliation{\Cem}
\author{Zahir Muhammad}
\affiliation{\Beihang}
\author{Zia~Ur~Rehman}
\affiliation{\Pakistan}
\author{Weisheng Zhao}
\affiliation{\Beihang}
\author{Adam Babiński}
\affiliation{\FUW}
\author{Maciej R. Molas}
\email{Maciej.Molas@fuw.edu.pl}
\affiliation{\FUW}

\begin{abstract}
Charge-density waves (CDWs) are ordered quantum states of conduction electrons accompanied by periodic lattice distortions. 
Raman scattering (RS) spectroscopy is therefore well suited for probing CDW-induced structural modulations. 
We investigate the CDW state in quasi-one-dimensional NbTe$_4$ using RS spectroscopy. 
At $T$=5~K, the resonantly enhanced Raman spectrum exhibits 25 phonon modes. Polarization-dependent measurements reveal a strong coupling between phonon-mode symmetry and crystallographic symmetry, with modes polarized parallel or perpendicular to the crystallographic $c$-axis, along which the one-dimensional structure is elongated. 
Temperature-dependent RS measurements identify a transition between commensurate and incommensurate CDW phases, accompanied by pronounced thermal hysteresis, with transition temperatures of approximately 45~K upon cooling and 90~K upon warming. 
The hysteresis width depends on the warming rate, indicating a finite nucleation rate of CDW domains and suggesting potential relevance for memory-device applications.

\end{abstract}

\maketitle


Two-dimensional (2D) van der Waals (vdW) materials, including graphene and transition metal dichalcogenides, exhibit a wide range of emergent physical properties.\cite{Geim2007, Koperski2017}
Among those phenomena, charge density waves (CDWs) have attracted considerable attention.\cite{Gruner1988,Yoshida2015,Lacinska2022}
The CDW is an ordered quantum state of conduction electrons forming a macroscopic standing-wave modulation of the charge density.\cite{Wilson1974, Gruner1988, Thorne1996}
This electronic ordering is accompanied by a periodic distortion of the crystal lattice, which modifies the lattice periodicity and lowers the total energy of the system.\cite{Wilson1974, Gruner1988, Thorne1996} The energetic cost of the lattice distortion is compensated by a reduction in the electronic energy, leading to a lowering of the Fermi level and the opening of an energy gap.\cite{Wilson1974,Gruner1988,Thorne1996}
CDWs are commonly realized in linear chain compounds and layered crystals.\cite{Bennett1999,Yoshida2014,Lin2020,Lacinska2022}
These materials exhibit a variety of unusual properties, including nonlinear electrical transport, enabling potential applications as memristive phase-switching nanodevices.\cite{Yoshida2014,Yoshida2015,LUDWICZAK2020}
Furthermore, many CDW systems are known to develop superconductivity under applied pressure.\cite{Ritschel2013,Yang2018}.

CDW materials exhibit at least two distinct phases, differing in the size and shape of domains containing periodic lattice distortions: the incommensurate CDW (ICDW) and the commensurate CDW (CCDW).\cite{Yoshida2014, Yoshida2015, Lacinska2022}. 
The ICDW state is characterized by a mismatch between the period of the modulated electronic charge density and the periodicity of the underlying atomic lattice.
With decreasing temperature of such a system, the emerging periodic lattice distortions spread uniformly over the entire crystal, making the periodicity of the electronic density commensurate with that of the atomic lattice, a state referred to as the CCDW.
The transition from the incommensurate state to the commensurate state is known as the lock-in transition, where the system "locks" its CDW modulation to the underlying lattice periodicity.\cite{Wilson1974, Gruner1988, Thorne1996}. 
Between the ICDW and CCDW phases, intermediate states may arise, characterized by locally commensurate domains separated by domain-wall phase slips (discommensurations), which maintain the average incommensurate periodicity.\cite{Thomson1994}
The periodic lattice distortion pattern accompanying a CDW is reported for several 2D materials.\cite{Yoshida2014, Yoshida2015, Lacinska2022}
In 1T-TaS$_2$, distorted lattices form David-star clusters.\cite{Yoshida2014, Yoshida2015, Lacinska2022}
The temperature of the transition depends on the direction of the process, $i.e.$ cooling or heating, thus we observe a hysteresis.\cite{Eaglesham1985, Mahy1986, Yoshida2014, Yoshida2015, Shi2017, Wang2018, Ramos2019, LUDWICZAK2020}.

In this paper, we investigate the bulk NbTe$_4$ using polarization-resolved Raman scattering (RS) under resonant excitation at 1.58~eV over a wide temperature range (5--300~K). 
Using a high-resolution experimental setup, we identify 25 phonon modes at 5~K. 
The observed polarization of the phonon modes, oriented parallel or perpendicular to the crystallographic axis, reflects their strong sensitivity to the crystal symmetry. 
Across the ICDW--CCDW transition, the RS spectra exhibit abrupt changes in phonon modes, including their disappearance, emergence, and energy shifts. 
Temperature cycling between 5~K and 300~K reveals a pronounced hysteresis in the Raman shifts of the phonon modes. 
Upon warming, the transition occurs at approximately 90~K, whereas upon cooling it is observed near 45~K. 
Furthermore, the transition temperature depends on the warming rate, with faster heating shifting the transition to higher temperatures, indicating finite relaxation kinetics of CDW domains. 
We propose that the phase transition corresponds to a structural transformation between the $P4/ncc$ and $P4/mcc$ phases, with the observed hysteresis arising from the finite nucleation speed of CDW domains.


\begin{figure*}[t]
    \centering
    \includegraphics[width=1\linewidth]{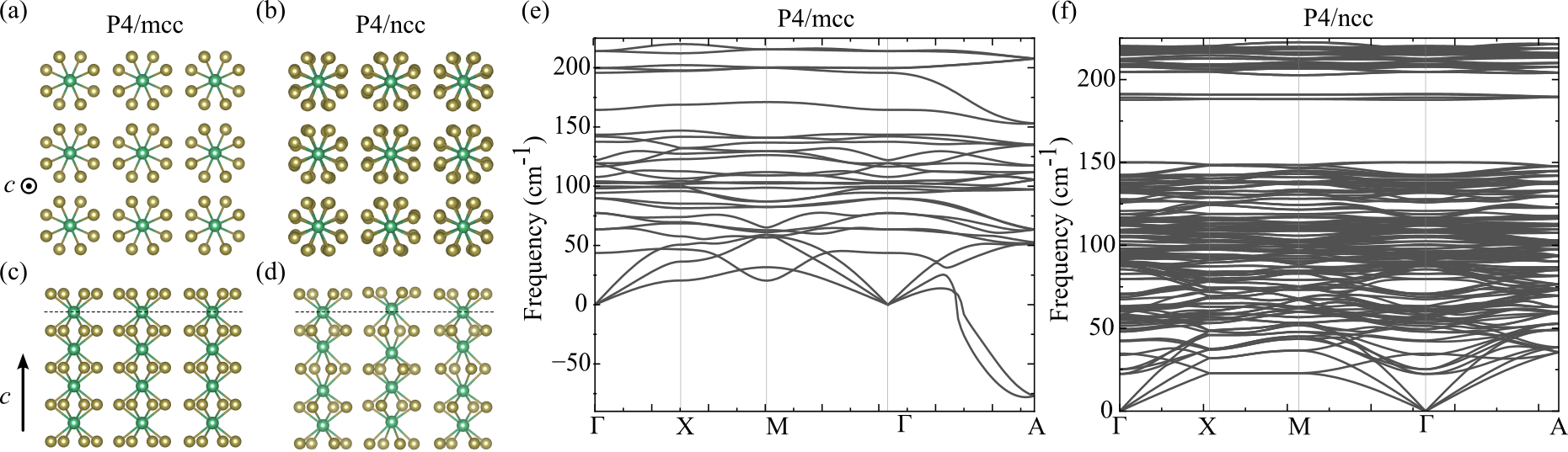}
    \caption
    {\label{fig1c}
    The crystal structures of NbTe$_4$ in the $P4/mcc$ and $P4/ncc$ phases. Green (yellow) spheres represent Nb (Te) atoms. Panels (a) and (b) show views along the crystallographic $c$ axis, while panels (c) and (d) show views perpendicular to the $c$ axis. In panels (c) and (d), horizontal dashed lines are included to more clearly highlight the displacement of Nb atoms in the $P4/ncc$ structure. Panels (e) and (f) shows the calculated phonon dispersion of NbTe$_4$ in the $P4/mcc$ and $P4/ncc$ space-group symmetry, respectively.}
\end{figure*}

The quasi-one-dimensional (quasi-1D) CDW material NbTe$_4$, is a member of the transition-metal tetratelluride family (MTe$_4$, M = transition metal).\cite{Boswell1983, Eaglesham1985, Mahy1986, Ikari1987, Tadaki1990, Bennett1999}
NbTe$_4$ crystallizes in a tetragonal structure with space group $P4/mcc$, schematically illustrated in Fig.~\ref{fig1c}(a, c).
The crystal structure consists of covalently bonded Nb--Te chains extending along the crystallographic $c$ axis.
Adjacent chains are coupled via interchain interactions that are weaker than covalent bonds but stronger than vdW forces, placing NbTe$_4$ in the class of quasi-1D materials.\cite{Bullett1984}
Single crystals of NbTe$_4$ were grown using chemical vapor transport technique, as presented in the Methods section of the Supplementary Information (SI)).
The crystals form needle-like morphologies elongated along the $c$ axis, and their room-temperature crystal structure ($T$=300~K) was confirmed by x-ray diffraction (XRD), as shown and discussed in Section~S1 of the SI.

At room temperature, NbTe$_4$ crystallizes in the $P4/mcc$ phase, which hosts an ICDW.
However, phonon dispersion calculations for that phase, shown in Fig.~\ref{fig1c}(e), reveal a lattice instability manifested by imaginary phonon modes (negative frequencies).
As a result, the system lowers its energy by adopting a stable $P4/ncc$ structure, which is fully relaxed and structurally modulated with respect to the parent $P4/mcc$ phase.
Fig.~\ref{fig1c}(f) displays the calculated phonon dispersion in the P4/ncc space group. Consistent with the structural stability of this phase, the phonon spectrum shows no imaginary phonon modes.
The crystal structures of $P4/mcc$ and $P4/ncc$ phases are shown in Fig.~\ref{fig1c}(a-d). 
In all panels, green and yellow spheres denote Nb and Te atoms, respectively. Panels (a) and (b) present views along the crystallographic $c$ axis, while panels (c) and (d) show views perpendicular to the $c$ axis. The $P4/ncc$ structure is modulated with respect to the $P4/mcc$ structure, giving rise to two primary structural differences. First, as shown in panel (b), the Te atoms within an individual chain are no longer aligned along the $c$ axis in the $P4/ncc$ structure, in contrast to the collinear arrangement observed in the $P4/mcc$ phase. Second, panel (d) illustrates that the Nb atoms in neighboring chains do not lie within the same plane in the $P4/ncc$ structure, unlike in the $P4/mcc$ phase. Horizontal dashed lines are included to emphasize the relative displacements of the Nb atoms.
This modulation of lattice induces a trimerization of the Nb and Te atoms along the quasi-1D chains, leading to an A--B--A trimer stacking pattern between neighboring chains.
Such a characteristic chain arrangement gives rise to the superstructure observed by scanning tunneling microscopy at 1.7~K.\cite{Galvis2023}
In contrast to the ICDW state of the $P4/mcc$ phase, the $P4/ncc$ structure supports a CCDW.
The lock-in transition from the ICDW to the CCDW phase occurs at $T\approx$~50~K, as evidenced by resistivity measurements\cite{Ikari1987, Tadaki1990, Yang2018, Galvis2023} and electron diffraction studies.\cite{Boswell1983, Eaglesham1985, Mahy1986}
At temperatures above 50~K, CDW discommensurations are observed.\cite{Boswell1983, Eaglesham1985, Mahy1986, Galvis2023}

In the ICDW phase, NbTe$_4$ crystallizes in the $P4/mcc$ (No.~124) space group and exhibits 30 phonon branches.
Group-theoretical analysis yields A$_\textrm{1g}$, A$_\textrm{2g}$, B$_\textrm{1g}$, B$_\textrm{2g}$, and E$_\textrm{g}$ Raman-active modes, as well as infrared-active modes A$_\textrm{1u}$, A$_\textrm{2u}$, B$_\textrm{1u}$, B$_\textrm{2u}$, and E$_\textrm{u}$.
Based on the polarization selection rules for this space group, 2~A$_\textrm{1g}$, 2~B$_\textrm{1g}$, and 6~E$_\textrm{g}$ Raman-active phonon modes are expected to be observed in the backscattering geometry with excitation and detection perpendicular to the crystallographic $c$ axis, as employed in our experiment.
In contrast, in the CCDW phase  adopts the $P4/ncc$ (No.~130) space group, resulting in a substantially enlarged unit cell with 180 phonon branches.
The corresponding irreducible representations include the same sets of Raman-active (A$_\textrm{1g}$, A$_\textrm{2g}$, B$_\textrm{1g}$, B$_\textrm{2g}$, E$_\textrm{g}$) and infrared-active (A$_\textrm{1u}$, A$_\textrm{2u}$, B$_\textrm{1u}$, B$_\textrm{2u}$, and E$_\textrm{u}$) modes.
According to the selection rules for the same backscattering geometry, 12~A$_\textrm{1g}$, 8~B$_\textrm{1g}$, and 44~E$_\textrm{g}$ modes are Raman active in the CCDW phase.



\begin{figure*}[t]
    \centering
    \includegraphics[width=1\linewidth]{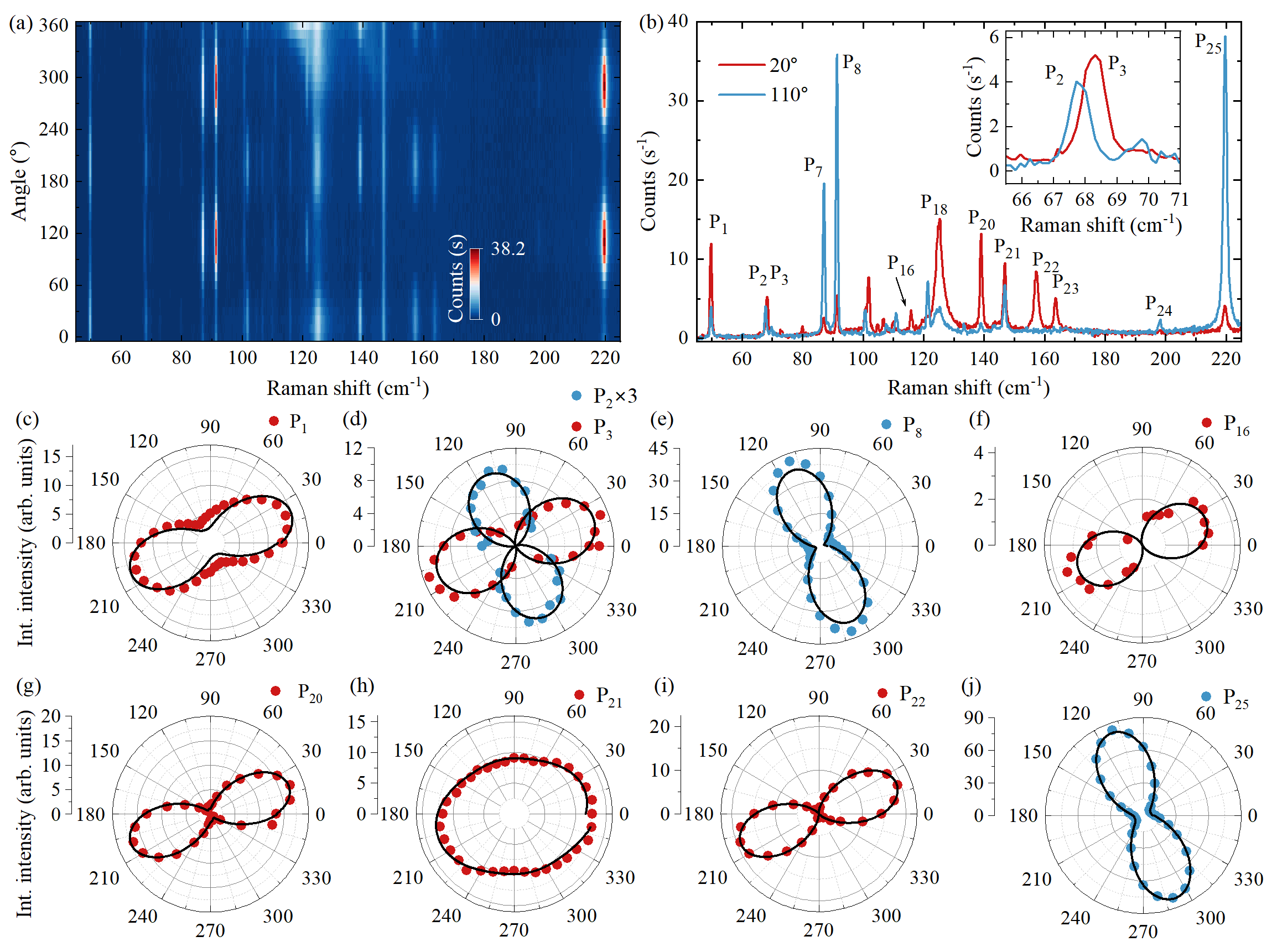}
    \caption
    {\label{fig2}
     (a) False-color map of low-temperature ($T$=5~K) Raman scattering (RS) spectra of the NbTe$_4$ crystal in the co-linear configuration, shown as a function of the angle of linear polarization with respect to the crystallographic orientation of the material, under 1.58~eV laser light excitation. 
     (b) RS spectra at two orthogonal detection angles, 20~$^{\circ}$ and 110~$^{\circ}$, extracted from the map in panel (a).
     The inset shows the RS spectrum around 68~cm$^{-1}$, where phonon modes P$_2$ and P$_3$ are present.
     (c)-(j) Polar plots of the integrated intensities of phonon modes P$_1$, P$_2$ and P$_3$, P$_8$, P$_{16}$, P$_{20}$, P$_{21}$, P$_{22}$, and P$_{25}$, respectively. For clarity, the integrated intensity of phonon mode P$_{2}$ is multiplied by 3 in panel (d).}
\end{figure*}

To verify theoretical predictions of phonon properties in the CDW phases, we performed polarization-resolved RS measurements on NbTe$_4$ in a co-linear configuration at low temperature ($T$=5~K).
The experimental details are provided in the Methods section of the SI.
In that configuration, the excitation and detection polarization axes are parallel, enabling symmetry analysis of the phonon modes.\cite{Zawadzka2021, BURUIANA2022, Antoniazzi_2023}
The polarization dependence of the RS signal with respect to the crystallographic orientation of NbTe$_4$ was probed by simultaneous rotation of the excitation and detection polarization axes.
A 785~nm (1.58~eV) excitation was selected, as it reveals the largest number of phonon modes in the spectrum, as demonstrated in Sec.~S2 of the SI.
A false-color map of the polarization-resolved RS spectra acquired at 5~K is shown in Fig.~\ref{fig2}(a).
All observed phonon peaks exhibit linear polarization, with their intensities maximized along one of two directions, at approximately 20$^{\circ}$ or 110$^{\circ}$.

\begin{figure*}[!t]
    \centering
    \includegraphics[width=1\linewidth]{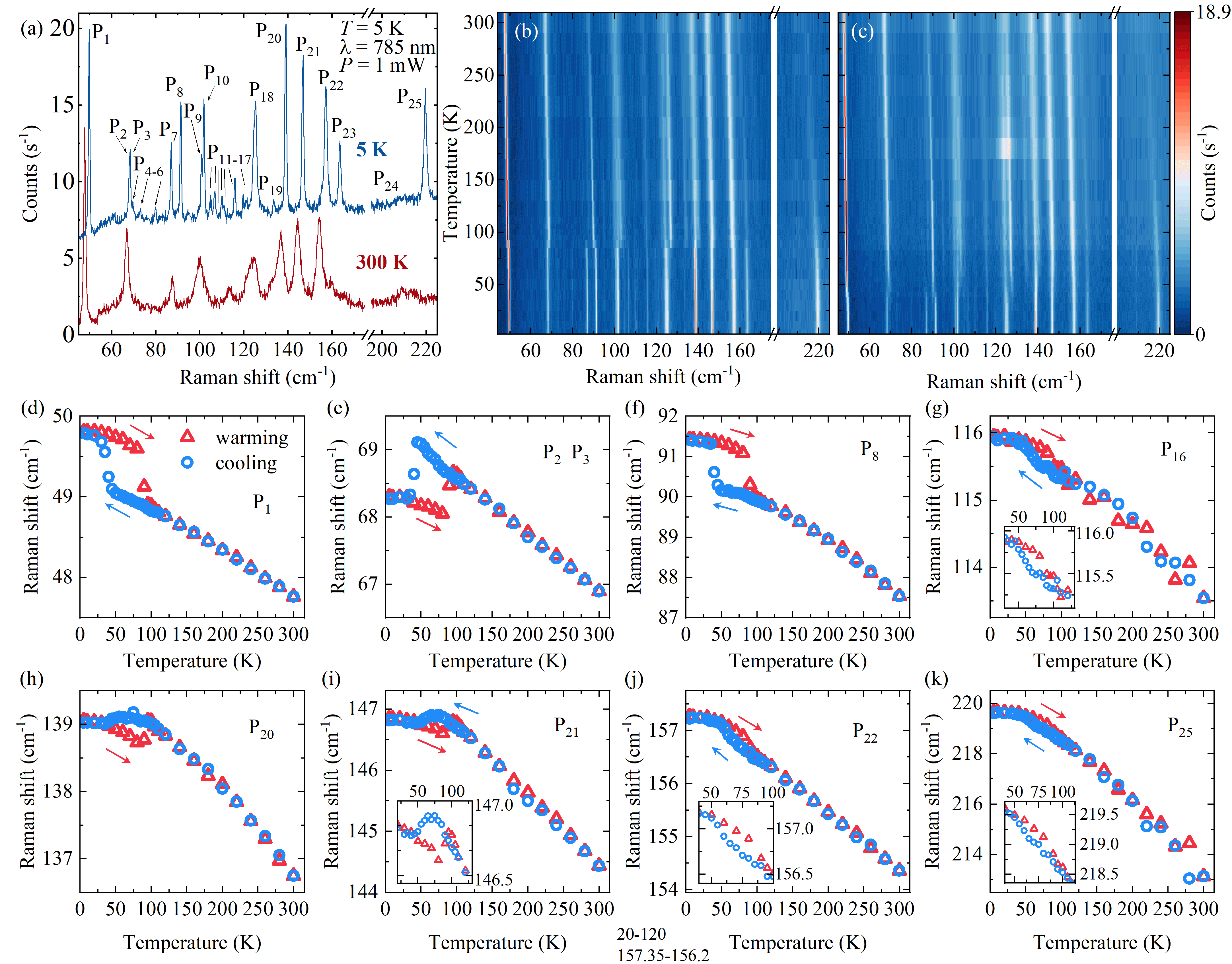}
    \caption
    {\label{fig3}
     (a) RS spectra of NbTe$_4$ under 785~nm (1.58~eV) excitation with 1~mW laser power at low ($T$~=~5~K) and room ($T$~=~300~K) temperatures. The spectra are vertically shifted for clarity. False-color maps of RS spectra as a function of temperature during warming from 5~K to 300~K (b) and cooling from 300~K to 5~K (c). (d-k) Temperature evolution of the Raman shifts during warming and cooling for the P$_1$, P$_2$ and P$_3$, P$_8$, P$_{16}$, P$_{20}$, P$_{21}$, P$_{22}$, P$_{25}$ phonon modes, respectively. The isnsets in panels (j, k) focus on the hysteresis.}
\end{figure*}

Representative RS spectra measured at these orthogonal polarization angles are presented in Fig.~\ref{fig2}(b).
In total, 25 Raman-active phonon peaks are resolved experimentally, whereas group-theoretical analysis predicts 64 phonon modes to be active in this backscattering geometry.
This indicates that certain Raman modes intensities are either below the detection limit or overlap with neighboring modes as a result of the finite spectral resolution of the experimental setup. This interpretation is supported by the phonon dispersion shown in Fig.~\ref{fig1c}(f).
Based on their polarization directions, the observed phonon modes can be divided into two groups (see Section~S3 of the SI).
The degree of polarization, defined by the intensity contrast between the two orthogonal angles, varies significantly among the modes.
For example, mode P$_{21}$ is weakly polarized, modes P$_1$, P$_8$, and P$_{18}$ show intermediate polarization, whereas modes P$_{16}$, P$_{22}$, and P$_{23}$ are fully polarized.
The inset of Fig.~\ref{fig2}(b) highlights the spectral range containing modes P$_2$ and P$_3$, which are fully polarized along orthogonal directions and separated by less than 1~cm$^{-1}$ in energy.

For quantitative analysis, integrated intensities of selected phonon modes was extracted and displayed as polar plots in Figs.~\ref{fig2}(c)–(j).
The angular dependence of the mode intensity, $I(\theta)$, is well described by
$I(\theta) = \left(|A| \sin^2(\theta - \phi) + |C| \cos(\alpha) \cos^2(\theta - \phi)\right)^2 + |C|^2 \sin^2(\alpha) \cos^4(\theta - \phi)$,
where $A$ and $C$ denote the amplitudes of the phonon modes, $\phi$ is the phase of the polarization dependence, and $\alpha$ is the phase difference.\cite{Ribeiro2015}
The polar plots reveal a clear 2-fold symmetry with a periodicity of 180$^{\circ}$ for all analyzed phonon modes.
The P$_1$, P$_3$, P$_{16}$, P$_{20}$, P$_{21}$, and P$_{22}$  modes are polarized along approximately 20$^{\circ}$, which is perpendicular to the crystallographic $c$ axis of the NbTe$_4$ crystal in the experimental geometry (see the sample photograph in Section~S3 of the SI).
In contrast, modes P$_2$, P$_8$, and P$_{25}$ are oriented along 110$^{\circ}$, corresponding to polarization parallel to the $c$ axis.
Similar polarization behavior of phonon modes has been reported in other anisotropic vdW materials.\cite{Zawadzka2021, BURUIANA2022, Zahir2024}
A definitive assignment of the observed Raman peaks to individual phonon modes remains challenging, as only 25 modes are resolved experimentally, compared to 64 modes predicted by theory.
Complementary room-temperature Raman measurements, as well as measurements performed with light propagation along the $c$ axis, are discussed in detail in Section~S3 of the SI.


Figure~\ref{fig3}(a) shows the RS spectra of NbTe$_4$ measured at low ($T$=5~K) and room ($T$=300~K) temperatures, corresponding to the CCDW and ICDW phases, respectively, since the transition between these two phases occurs at $T\approx$50~K.\cite{Boswell1983, Eaglesham1985, Mahy1986, Ikari1987, Tadaki1990, Yang2018, Galvis2023}
At low temperature, 25 phonon modes are observed experimentally, whereas theory predicts a substantially larger number (64).
In contrast, at room temperature, 15 phonon modes are observed, while fewer Raman-active modes (10) are predicted by group-theoretical analysis.
This discrepancy suggests the activation of normally forbidden infrared or acoustic modes, which may arise from lattice distortions associated with CDW formation.
Such distortions can relax Raman selection rules through partial breaking of inversion symmetry, even if the structural distortions are small.
Consistent with this interpretation, several phonon modes observed in the Raman spectra of NbTe$_4$ have previously been assigned to infrared-active modes.\cite{Luo2024, Nataj2024, Shuang2024}
Similarly, the appearance of infrared-active phonon modes has been reported in TaTe$_4$, another CDW material with the same crystal structure.\cite{Nataj2024}

A comprehensive comparison between experimentally observed Raman peaks and theoretical predictions, including different crystal orientations with respect to the light propagation direction, is provided in Section~S3 of the SI.

We performed temperature-dependent RS measurements to trace the phase transition between the low-temperature CCDW phase with $P4/ncc$ symmetry and the high-temperature ICDW phase with $P4/mcc$ symmetry.
Resistivity and electron diffraction studies show a transition at approximately 50~K associated with a lock-in transition into the fully CCDW phase.\cite{Boswell1983, Eaglesham1985, Mahy1986, Ikari1987, Tadaki1990, Yang2018, Galvis2023}
In addition, anomalies in resistivity have been observed near 200~K, suggesting the presence of intermediate phases.\cite{Eaglesham1985, Mahy1986, Ikari1987, Tadaki1990, Yang2018, Galvis2023}
Previous studies have demonstrated that the transition temperature depends on the thermal history of the sample, with different values observed during cooling and warming cycles.\cite{Eaglesham1985, Mahy1986}
For example, the transition occurs at $\sim$50~K during cooling, whereas during warming it has been reported at $\sim$100~K\cite{Mahy1986} or even $\sim$150~K.\cite{Eaglesham1985}
In previous Raman studies of NbTe$_4$, this phase transition was not observed, likely because the measurements were limited to temperatures between 70 and 300~K, $i.e.$, above the reported transition temperature.\cite{Nataj2024, Luo2024}

The experiment was initiated at 5~K, and the RS spectra were subsequently measured as a function of temperature during warming from 5~K to 300~K.
Measurements were then repeated during cooling from 300~K back down to 5~K.
False-color maps showing the evolution of the RS spectra during warming and cooling are presented in Figs.~\ref{fig3}(b) and \ref{fig3}(c), respectively.
During warming, the phase transition begins at approximately 90~K.
Of the 25 phonon modes present in the low-temperature phase, only 15 remain above the transition, indicating that the complex crystal structure described by the $P4/ncc$ space group is replaced by a simpler structure with a smaller primitive unit cell described by the $P4/mcc$ space group.
During cooling, the lock-in transition starts at about 45~K and is completed near 30~K, in good agreement with previous reports.\cite{Boswell1983, Eaglesham1985, Mahy1986, Ikari1987, Tadaki1990, Yang2018, Galvis2023}
Thermal hysteresis in NbTe$_4$ has previously been observed only in electron diffraction experiments.\cite{Eaglesham1985, Mahy1986}
Hysteresis in the phase transition temperature detected by Raman spectroscopy has also been reported for other CDW materials, such as 1T-TaS$_2$.\cite{Shi2017, Wang2018, Ramos2019}
At higher temperatures, no anomalies indicative of intermediate phase transitions are observed in the RS spectra, in contrast to earlier reports based on resistivity measurements.\cite{Eaglesham1985, Mahy1986, Ikari1987, Tadaki1990, Yang2018, Galvis2023}
Figures~\ref{fig3}(d)–(k) show the temperature evolution of the Raman shifts during warming and cooling for selected phonon modes P$_1$, P$_2$, P$_3$, P$_8$, P$_{16}$, P$_{20}$, P$_{21}$, P$_{22}$, and P$_{25}$.
The width of the hysteresis in the Raman shift is approximately 60~K.
The hysteresis is most pronounced for modes P$_1$, P$_2$, P$_3$, P$_8$, and P$_{20}$, whereas it is less apparent for other modes that do not exhibit abrupt energy changes at the phase transition.
The modes P$_1$ and P$_8$, presented in Figs.~\ref{fig3}(d) and (f), exhibit distinct shifts to lower energies during warming, while the mode P${_20}$, shown in panel (h) of the figure, shifts toward higher energies.
The modes P$_2$ and P$_3$, shown together in Fig.~\ref{fig3}(e) and separated by less than 1~cm$^{-1}$, show a distinct behavior: the mode P$_2$ is visible at 5~K but quenches after warming, while the higher-energy mode P$_3$ simultaneously gains intensity.


\begin{figure}[!t]
    \centering
    \includegraphics[width=1\linewidth]{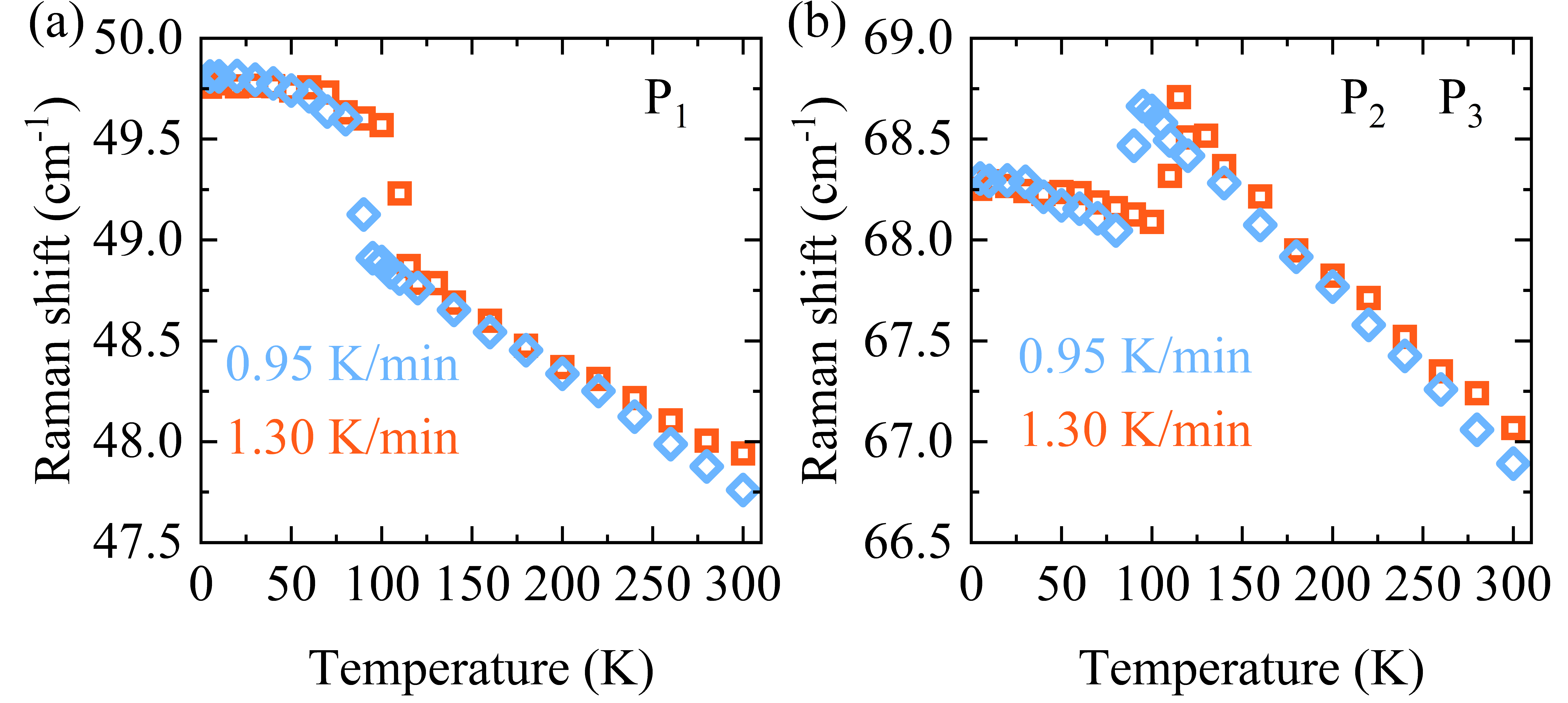}
    \caption
    {\label{fig4}
     Temperature evolution of the Raman shifts with warming rates 1.30~K/min and 0.95~K/min for P$_1$ at panel (a) and  P$_2$, P$_3$ at panel (b). }
\end{figure}

To further elucidate the kinetics of the phase transition in NbTe$_4$, we investigated the influence of the warming rate on the transition temperature.
RS spectra were measured as a function of temperature from 5 to 300~K using warming rates of 0.95 and 1.30~K/min.
Figures~\ref{fig4}(a) and \ref{fig4}(b) show the temperature evolution of the Raman shifts of phonon modes P$_1$, P$_2$, and P$_3$ for the two warming rates, revealing clear differences in their behavior.
For the slower warming rate (0.95~K/min), the phase transition occurs at approximately 90~K, whereas for the faster warming rate (1.30~K/min) the transition shifts to a higher temperature of about 110~K.
This behavior suggests a finite nucleation speed of CDW domains.
Similar warming-rate-dependent shifts of the transition temperature have been reported for other CDW materials, such as 1T-TaS$_2$.\cite{Yoshida2014, Yoshida2015}
The warming rate also affects the width of the thermal hysteresis: slower warming results in a narrower hysteresis, while faster warming leads to a broader one.
This property may be relevant in the context of memory-device applications.

In summary, we investigated the CDW phases in NbTe$_4$ using Raman scattering.
Under resonant laser excitation, 25 phonon modes are observed at 5~K, exceeding those previously reported. 
Polarization-resolved RS measurements allow determination of phonon-mode symmetries, revealing a strong sensitivity to crystal symmetry, with modes polarized either parallel or perpendicular to the crystallographic $c$ axis. 
Temperature-dependent RS measurements performed during the warming from 5~K to 300~K and subsequent cooling back to 5~K enable direct probing of the phase transition between the ICDW and CCDW phases.
The transition is accompanied by an abrupt appearance, disappearance, and energy shifts of phonon modes, as well as a pronounced hysteresis in the Raman shifts.
The transition occurs at approximately 90~K upon warming and 45~K upon cooling, consistent with a structural change between the $P4/ncc$ and $P4/mcc$ phases. 
Furthermore, the transition temperature depends on the warming rate, with faster heating shifting the transition to higher temperatures, indicating finite relaxation kinetics and a finite nucleation rate of CDW domains. 
These properties of NbTe may be promising for the development of memory-device technologies.

\section*{Acknowledgment}
The work was supported by the National Science Centre, Poland (Grant No. 2024/53/N/ST3/02638).

\bibliographystyle{apsrev4-2}
\bibliography{biblio}

\newpage
\onecolumngrid

\include{ESI_arxiv}

\end{document}

%% file: ESI_arxiv.tex
\renewcommand{\thefigure}{S\arabic{section}.\arabic{figure}}
\renewcommand{\thesection}{S\arabic{section}}

\begin{center}
	{\large{ {\bf Supplemental Material: \\ Raman scattering fingerprints of the charge density wave state in one-dimensional NbTe$_4$}}}
	\vskip0.5\baselineskip{Natalia Zawadzka,{$^{1}$} Cem Sevik,{$^{2}$} Zahir Muhammad,{$^{3}$} Zia~Ur~Rehman,{$^{4}$} Weisheng Zhao,{$^{3}$} Adam Babiński,{$^{1}$} \\ and Maciej R. Molas{$^{1}$}}

	\vskip0.5\baselineskip{\em$^{1}$ University of Warsaw, Faculty of Physics, 02-093 Warsaw, Poland \\$^{2}$ Department of Physics-NANOlight Center of Excellence, University of Antwerp, Antwerp B-2020, Belgium \\$^{3}$ National Key Lab of Spintronics, International Innovation Institute, Beihang University, Hangzhou 311115, China \\$^{4}$ Nanoscale Synthesis \& Research Laboratory, Department of Applied Physics, University of Karachi, Karachi 75270, Pakistan}
	\end{center}

This Supporting Information provides: \ref{Methods} - Details of the crystal growth, X-ray difraction, Raman scattering, and first principles calculations. \ref{S1} - Room-temperature X-ray diffraction analysis of NbTe$_4$. \ref{S2} - Crystal structure and phonon dispersion of NbTe$_4$. \ref{S3} - Excitation-dependent Raman scattering study of the NbTe$_4$.

\renewcommand{\thesection}{Methods}
\section{\label{Methods}}

\noindent
\textbf{Crystal growth.}\\

Single crystals of NbTe$_4$ were synthesized using the two-zone chemical vapor transport (CVT) technique. 
Niobium powder (99.99$\%$ purity) and tellurium powder (99.99$\%$ purity) were mixed in a weight ratio of 1:4 and sealed in a quartz tube under a vacuum of $10^{-6}$ Torr, with iodine added as the transport agent. 
The reaction zone was maintained at 1273 K and the growth zone at 1203 K. 
The furnace was slowly heated to the target temperatures over 48 hours and kept at these conditions for one week. 
Subsequently, both zones were cooled down to room temperature at a rate of 5~K per hour. Upon opening the furnace, high-quality single crystals of NbTe$_4$ were obtained.\\

\noindent
\textbf{X-ray diffraction.}\\
X-ray powder diffraction (XRD) was performed to confirm the crystal structure and phase purity of the synthesized NbTe$_4$ crystals. Measurements were carried out at room temperature using a SmartLab 3~kW diffractometer equipped with a Cu anode source. The crushed bulk crystalline powder was sealed in a 0.5~mm glass capillary and examined over a 2$\theta$ range of 5$^{\circ}$ to 90$^{\circ}$ with Cu K$\alpha$ radiation ($\lambda$ = 0.15418~nm, 8.04~keV). The obtained diffraction pattern verified the expected structure and indicated a phase-pure material.
\\
\noindent

\noindent
\textbf{Raman scattering measurements.}\\
Raman scattering spectra were measured under laser excitation of $\lambda$=785~nm (1.58~eV). 
The excitation light in those experiments was focused by means of a 50x long-working distance objective with a 0.55 numerical aperture (NA) producing a spot of about 1 $\mu$m diameter. 
The signal was collected via the same microscope objective (the backscattering geometry), sent through a 0.75 m monochromator, and then detected using a liquid nitrogen-cooled charge-coupled device (CCD) camera. 
All measurements were performed with the samples placed on a cold finger in a continuous-flow cryostat mounted on x-y manual positioners. 
The excitation power focused on the sample was kept at 1~mW during all measurements. 
The polarization-resolved RS measurements were performed in the co-linear (XX) and cross-linear (XY) configurations, which correspond to the parallel orientation of the excitation and detection polarization axes. 
The analysis of the RS signal was done using a motorized half-wave plate mounted on top of the microscope objective, which provides simultaneous rotation of the polarization axis on excitation and detection.
\\
\noindent

\noindent
\textbf{Theoretical calculations.}\\
Structural relaxations and force calculations were carried out within the density functional theory (DFT) using the Perdew–Burke–Ernzerhof (PBE) exchange–correlation functional of the generalized gradient approximation (GGA)~\cite{Kresse1993}, as implemented in VASP~\cite{Pedrew1996}. 
A plane-wave basis with an energy cutoff of 520 eV was employed throughout. 
Brillouin zone integrations for structural relaxations used $\Gamma$-centered Monkhorst–Pack~\cite{Monkhorst1976} meshes of 8$\times$8$\times$2 and 12$\times$1$\times$12 for the $P4/ncc1$ and $P4/mcc$ structures, respectively. 
Electronic self-consistency was converged to $10^{-7}$ eV, while ionic relaxations were stopped when forces fell below $10^{-2}$ eV/AA.

Phonon dispersions were computed using the finite displacement method implemented in Phonopy~\cite{Togo2015}. 
For second-order interatomic force constants, we employed a $2 \times 2 \times 1$ supercell with a $4 \times 4 \times 2$ $\Gamma$-centered \textit{k}-mesh for the $P4/ncc1$ phase, and a $2 \times 2 \times 2$ supercell with a $6 \times 6 \times 6$ $\Gamma$-centered \textit{k}-mesh for the $P4/mcc$ phase.

\clearpage
\newpage
\renewcommand{\thesection}{S\arabic{section}}
\setcounter{section}{0}
\setcounter{figure}{0}
\section{Room-temperature X-ray diffraction analysis of NbTe$_4$.\label{S1}}

\begin{figure}[!h]
    \centering
    \includegraphics[width=0.45\linewidth]{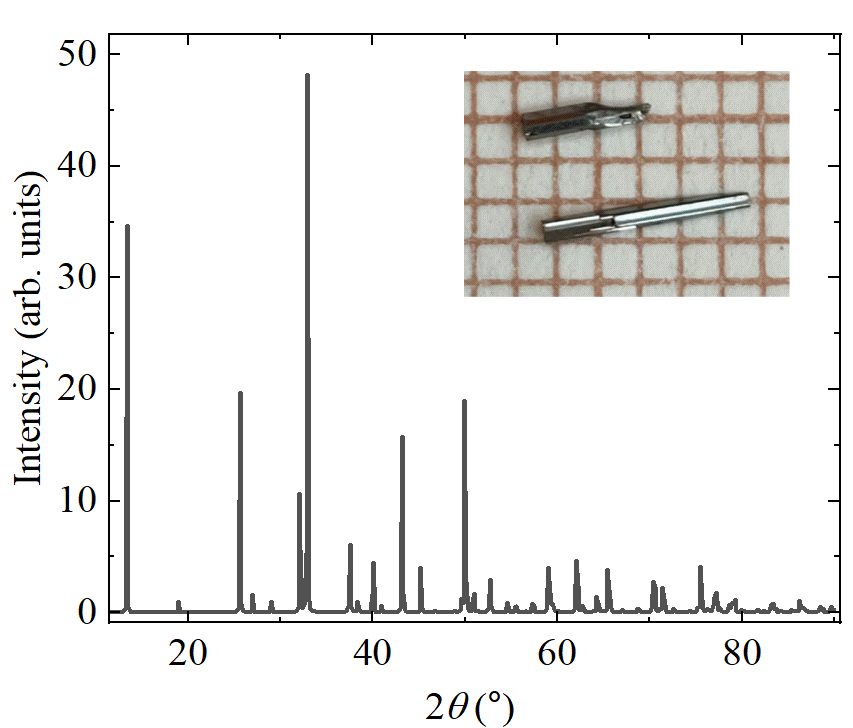}
    \caption
    {\label{SI_XRD}
    Powder X-ray diffraction (XRD) pattern of NbTe$_4$ sample. 
    The inset shows a photograph of the NbTe$_4$ single crystals.  }
\end{figure}

The $P4/mcc$ (Space Group No.~124) crystal structure of the investigated NbTe$_4$ sample at room temperature ($T$=300~K) was confirmed by powder X-ray diffraction.
The corresponding data are presented in Fig.~\ref{SI_XRD}.
The sharp diffraction peaks, with no detectable impurity phases, indicate the high crystalline quality of the sample.
The inset of Fig.~\ref{SI_XRD} displays an optical photograph of the NbTe$_4$ crystals, which exhibit a needle-like morphology characteristic of the P4/mcc structure and are elongated along the crystallographic $c$ axis.
These crystals were used for Raman scattering (RS) measurements.
The photograph was taken along the direction of the incident laser beam, confirming that the excitation laser beam was perpendicular to the crystallographic $c$ axis.

\clearpage
\newpage
\setcounter{figure}{0}
\section{Excitation-dependent Raman scattering study of the NbTe$_4$\label{S2}}
\begin{figure}[!h]
    \centering
    \includegraphics[width=0.65\linewidth]{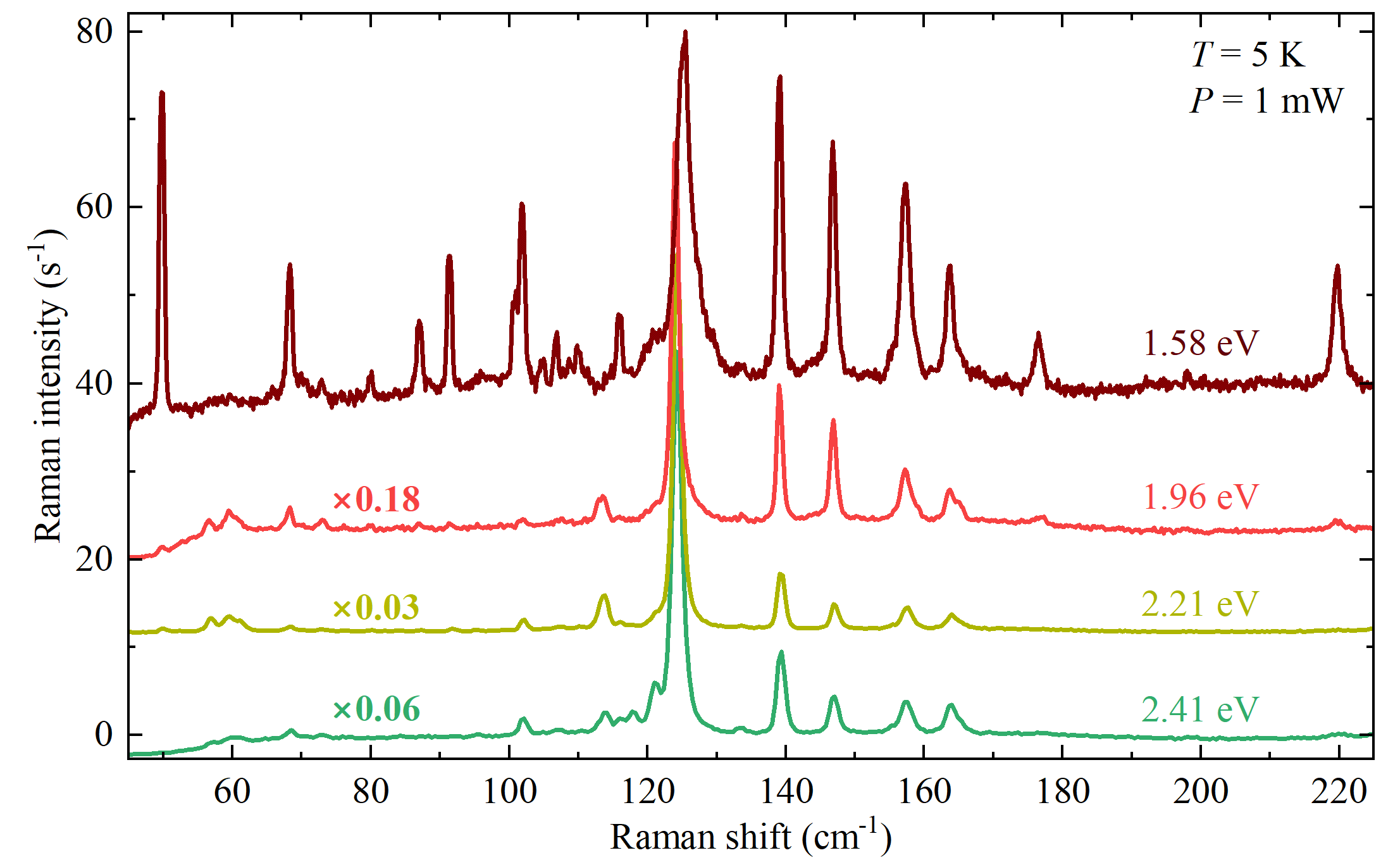}
    \caption
    {\label{SI_resonant_conditions}
    The Raman scattering spectra measured on the NbTe$_4$ crystal at 5~K under excitation energies: 2.41~eV, 2.21~eV, 1.96~eV, and 1.58~eV, using a laser power of 1~mW.
    Spectra are multiplied by scaling factor and vertically shifted for clarity. }
\end{figure}

To determine the optimal RS conditions, defined as those yielding the highest-quality Raman spectra, low-temperature ($T$=5~K) RS measurements were performed using four excitation energies (2.41~eV, 2.21~eV, 1.96~eV, and 1.58~eV), as shown in Fig.~\ref{SI_resonant_conditions}.
As shown in the figure, both the intensity and the spectral profile of the Raman response depend strongly on the excitation energy.
The Raman scattering intensity obtained with the 2.21~eV excitation is approximately 33 times larger than that obtained with the 1.58~eV excitation.
However, due to the substantially higher spectral resolution in the low-energy excitation range around 1.58~eV, which degrades with increasing excitation energy, a larger number of Raman peaks can be clearly resolved under 1.58~eV excitation compared with higher-energy excitations.
Consequently, despite its lower overall intensity, the 785~nm (1.58~eV) excitation wavelength was selected for all subsequent measurements, as it provides superior spectral resolution and enables a more detailed analysis of the vibrational modes.

\clearpage
\newpage
\setcounter{figure}{0}
\section{Angle-resolved Raman-based classification of phonon modes.\label{S3}}

\begin{figure}[!h]
    \centering
    \includegraphics[width=0.8\linewidth]{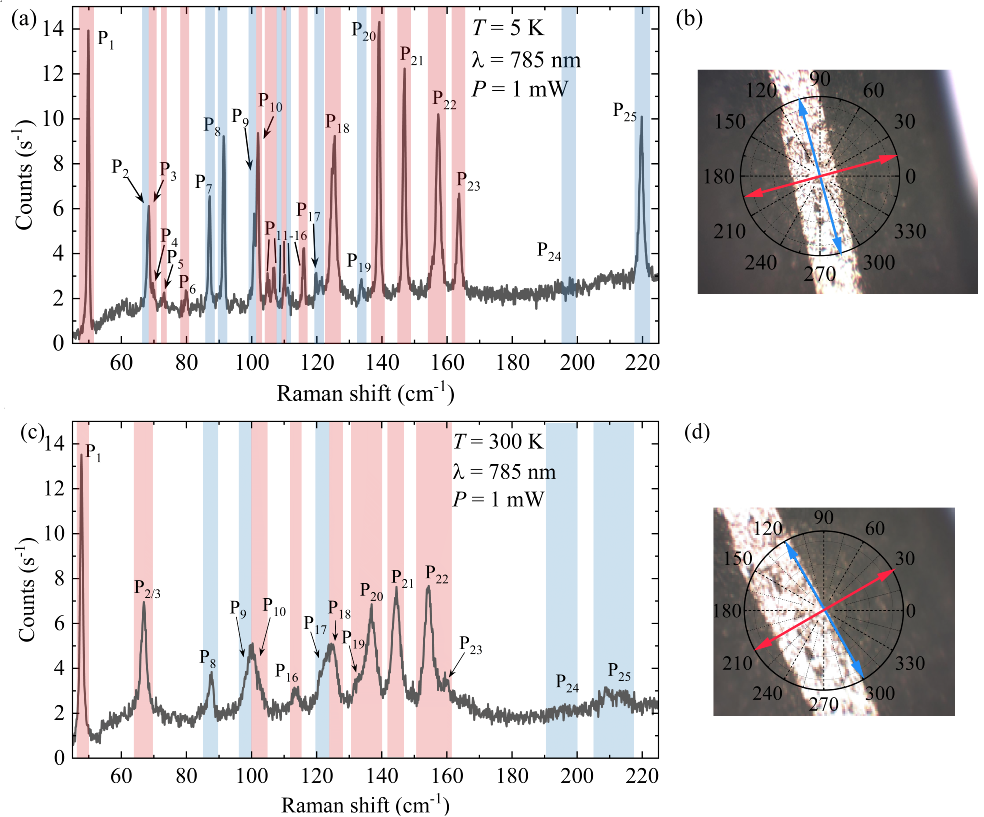}
    \caption
    {\label{SI_clasification}
RS spectra of NbTe$_4$ measured at 5~K (a) and 300~K (c) using 785~nm excitation and 1~mW power. Phonon modes polarized along the crystallographic $c$ axis are marked in blue, while modes polarized perpendicular to the $c$ axis are marked in red. (b, d) Photo of the NbTe$_4$ crystal investigated by RS, taken using a $5\times$ objective lens, corresponding to the measurements in panels (a) and (c), respectively. An overlaying polar plot shows the polarization directions corresponding to the blue (parallel to $c$) and red (perpendicular to $c$) marked phonon modes.  }
\end{figure}

Polarization-resolved RS measurements of the NbTe$_4$ crystal, performed in the co-linear (XX) configuration at low temperature ($T = 5\text{ K}$), revealed that the observed phonon modes are linearly polarized along one of two principal directions: 20~$^{\circ}$ or 110~$^{\circ}$. Fig.~\ref{SI_clasification}(a) presents the RS spectrum, where phonon modes linearly polarized along the 20~$^{\circ}$ angle are marked in red, and those polarized along the 110~$^{\circ}$ angle are marked in blue. Fig.~\ref{SI_clasification}(b) is a photo of the investigated sample with an overlaying polar plot. The blue and red arrows in the polar plot indicate the directions of polarization corresponding to the respective phonon modes.
It is evident that the direction 110~$^{\circ}$ aligns with the crystallographic $c$ axis, while the 20~$^{\circ}$ direction is polarized perpendicularly to the $c$ axis. 
We also provided polarization-resolved RS measurements at room temperature (300~K), with the crystal in a slightly different rotational position. The observed phonon modes are again divided into two groups based on their linear polarization: those polarized along the 120$^{\circ}$ direction, which aligns with the crystallographic $c$ axis, and those polarized along the 30$^{\circ}$ direction, which is perpendicular to the $c$ axis.
Fig.~\ref{SI_clasification}(c) presents the RS spectrum, where phonon modes polarized along the 30~$^{\circ}$ angle are marked in red, and those polarized along the 120~$^{\circ}$ angle are marked in blue. Fig.~\ref{SI_clasification}(d) is a photo of the investigated sample with an overlaying polar plot, where blue and red arrows indicate the corresponding directions of polarization. The group of phonon modes marked in blue is polarized along the crystallographic $c$ axis, and those marked in red are perpendicular to the $c$ axis. It is observed that the polarization directions of the phonon modes at both 5~K and 300~K, relative to the crystal orientation, remain consistent. This consistency indicates that the underlying symmetry and the alignment of the vibrational modes with respect to the crystallographic axes are preserved across this temperature range, despite the phase transition.

\begin{figure}[!h]
    \centering
    \includegraphics[width=0.65\linewidth]{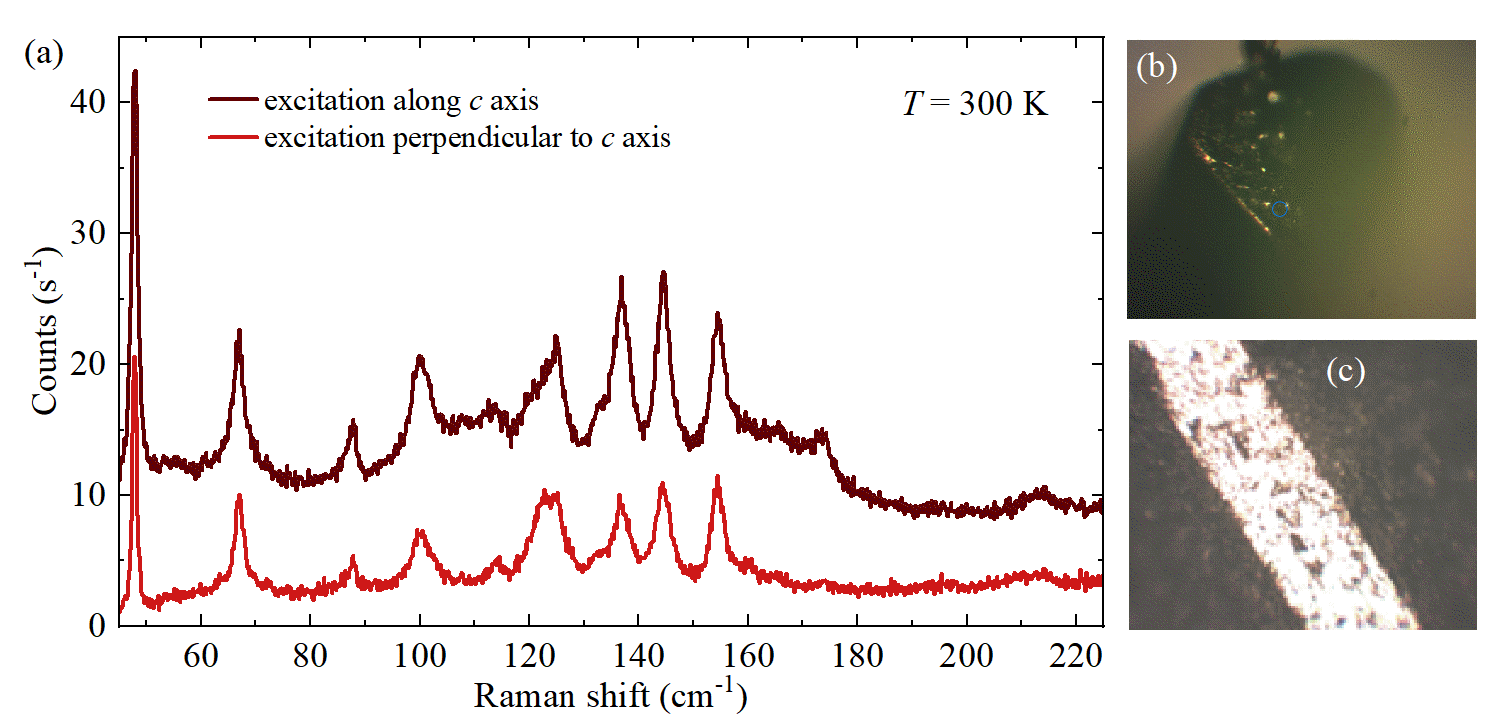}
    \caption
    {\label{SI_krysztal_vertykalny_vs_plaski}
   (a) RS spectra of NbTe$_4$ at 300~K measured with the incident laser excitation parallel and perpendicular to the crystallographic $c$ axis. The spectra are vertically shifted for clarity. (b, c) Photos of the crystals oriented in geometry when incident laser beam is parallel an perpendicular to crystallographic $c$ axis, respectively.}
\end{figure}

RS spectra were measured at 300 ~K using two distinct crystal orientations. The crystal was excited with the incident light polarized either parallel to the crystallographic $c$ axis (with the crystal oriented vertically) or perpendicular to it (with the crystal oriented horizontally).
The comparison of the two resulting spectra, alongside photographs illustrating the measurement geometry, is shown in Fig.~\ref{SI_krysztal_vertykalny_vs_plaski}. 

When the crystal is placed horizontally, excitation and detection are perpendicular to the crystallographic \textit{c} axis.
For the P4/mcc space group, 2~A$_{1g}$, 2~B$_{1g}$, and 6~E$_{g}$ Raman-active phonon modes are expected in the backscattering geometry.
Conversely, when the crystal is placed vertically, the excitation and detection are aligned  along the crystallographic \textit{c} axis. The 2~A$_{1g}$, 2~B$_{1g}$, and 2~B$_{2g}$ Raman-active phonon modes are expected. As the scattering geometry changes between these two configurations, certain phonon modes should disappear while others emerge in the spectra.
However, no significant difference is observed between the two spectra. Only a variation in the modes' intensity is apparent. This lack of dependence on the scattering geometry suggests that the observed phonon modes are likely due to resonant excitation.

\begin{figure}[!h]
    \centering
    \includegraphics[width=1\linewidth]{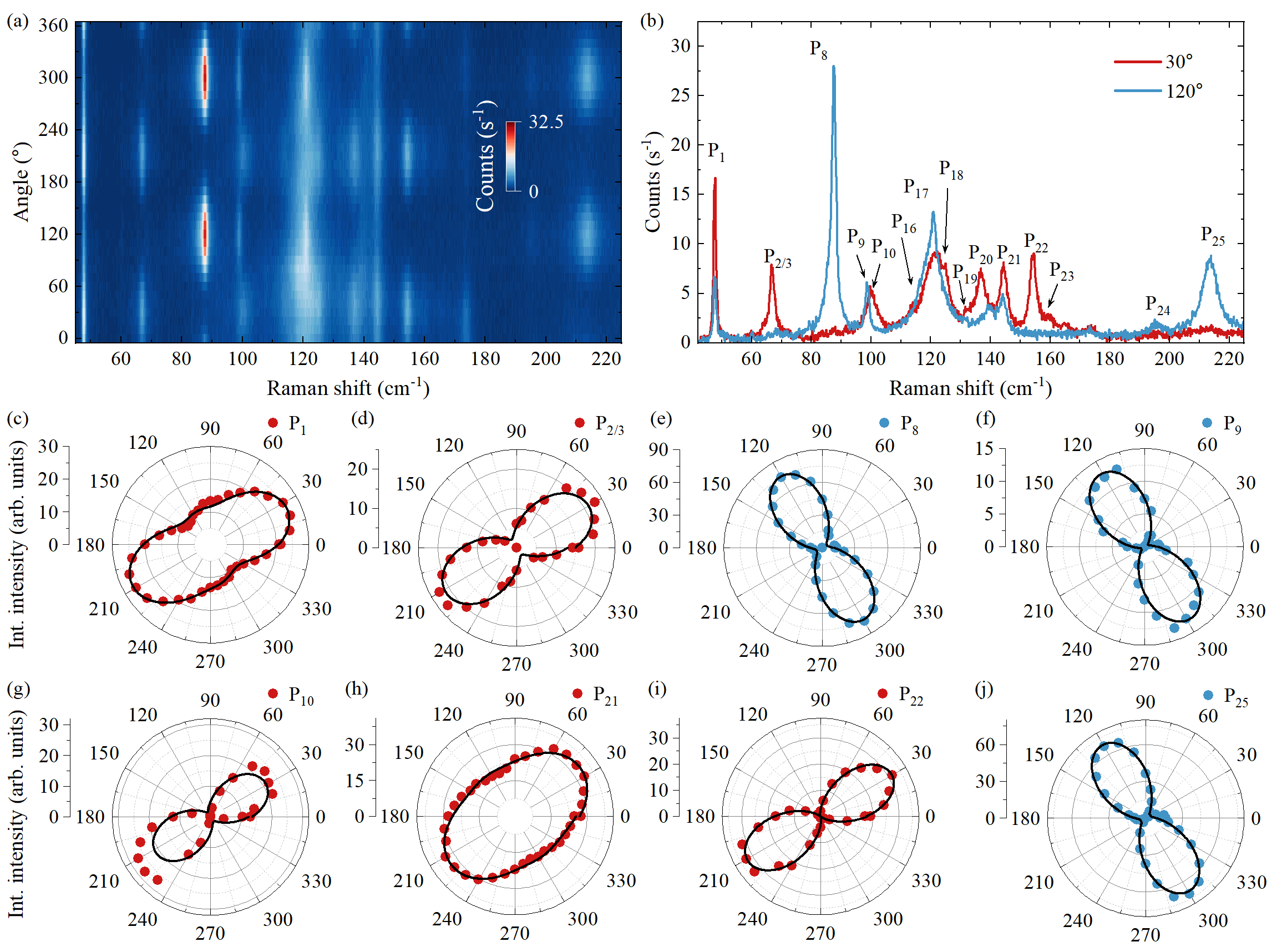}
    \caption
    {\label{SI_polaryzacja_plaski}
   (a) False-color map of RS spectra of the NbTe$_4$ crystal in the co-linear configuration, shown as a function of the angle of linear polarization, under 1.58~eV laser light excitation at 300~K. We excited the crystal perpendicular to the $c$ crystallographic axis. 
     (b) RS spectra at two orthogonal detection angles, 30~$^{\circ}$ and 120~$^{\circ}$, extracted from the map in panel (a).
     (c-j) Polar plots of the integrated intensities of phonon modes P$_1$, P$_{2/3}$, P$_8$, P$_{9}$, P$_{10}$, P$_{21}$, P$_{22}$, and P$_{25}$, respectively. }
\end{figure}

\begin{figure}[!h]
    \centering
    \includegraphics[width=1\linewidth]{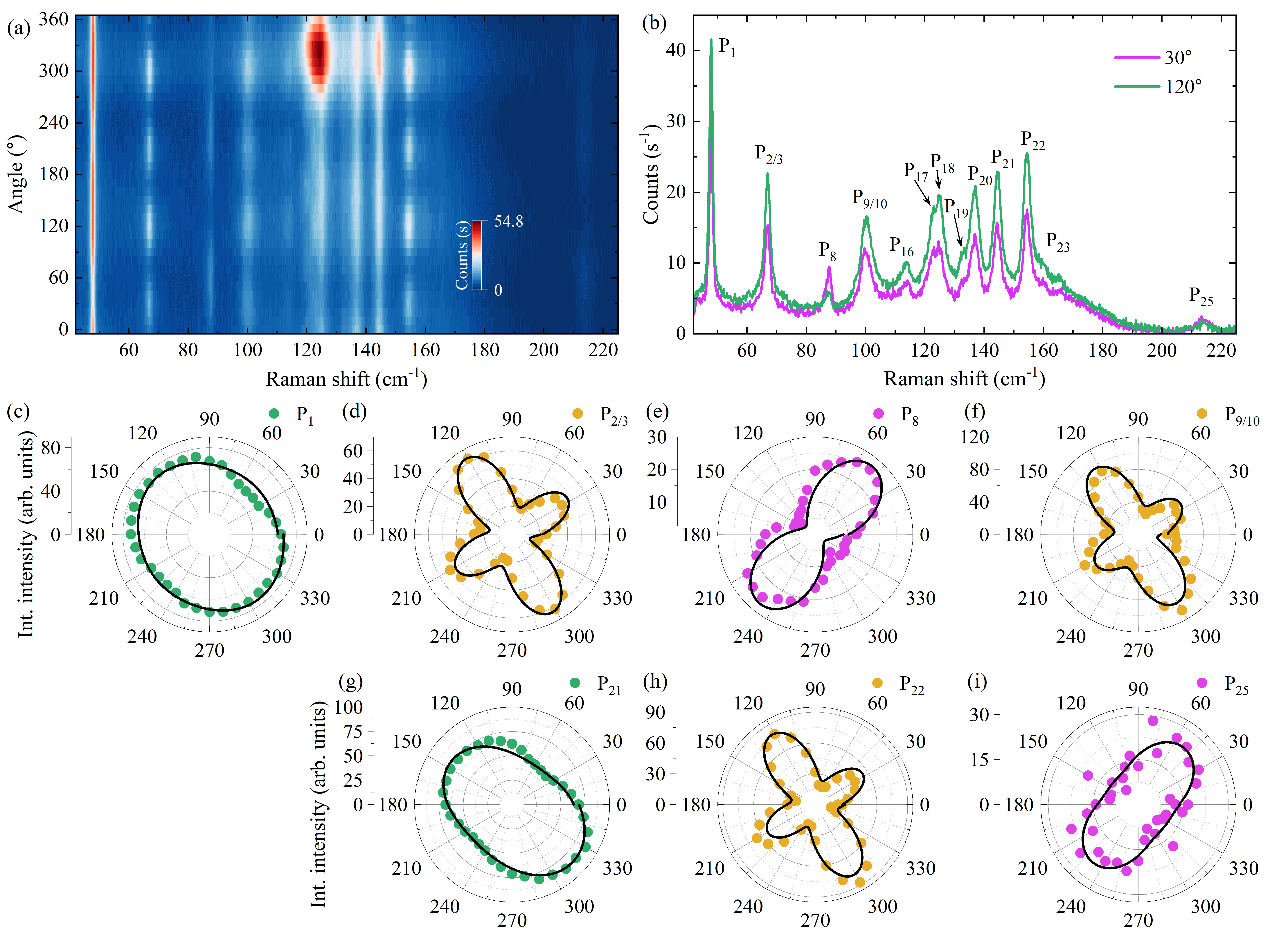}
    \caption
    {\label{SI_polaryzacja_wertykalny}
   (a) False-color map of RS spectra of the NbTe$_4$ crystal in the co-linear configuration, shown as a function of the angle of linear polarization, under 1.58~eV laser light excitation at 300~K. We excited the crystal along the $c$ crystallographic axis. 
     (b) RS spectra at two orthogonal detection angles, 30~$^{\circ}$ and 120~$^{\circ}$, extracted from the map in panel (a).
     (c-i) Polar plots of the integrated intensities of phonon modes P$_1$, P$_{2/3}$, P$_8$, P$_{9/10}$, P$_{21}$, P$_{22}$, and P$_{25}$, respectively.  }
\end{figure}

Polarization-resolved RS measurements were performed in the co-linear configuration for both crystal orientations (horizontal and vertical), as previously described. The results are presented in Figs.~\ref{SI_polaryzacja_plaski} and \ref{SI_polaryzacja_wertykalny}. 
In panels (a) of both figures, false-color maps present the RS as a function of the angle of linear polarization. Panels (b) show the corresponding RS spectra measured at two orthogonal detection angles: 30~$^{\circ}$ and 120~$^{\circ}$.
In panels (c-j) in Fig.~\ref{SI_polaryzacja_plaski} and (c-i) in Fig.~\ref{SI_polaryzacja_wertykalny} are polar plots of the integrated intensities of phonon modes P$_1$, P$_{2/3}$, P$_8$, P$_{9}$, P$_{10}$, P$_{21}$, P$_{22}$, and P$_{25}$, respectively. Note that in Fig.~\ref{SI_polaryzacja_wertykalny}(f) is presented mode P$_{9/10}$ as modes P$_{9}$ and P$_{10}$ formed one broader peak in crystal placed vertically.

As the shape of the spectra is the same in crystals placed horizontally and vertically the polarization properties of the phonon modes are different. There are three main differences: 

(1) With the horizontally placed crystal, all phonon modes display 2-fold symmetry with an angle period of 180~$^{\circ}$ and one of two orthogonal polarization axes; see polar plots on panels (c-j). When the crystal is placed vertically, the phonon modes display 2-fold symmetry with an angle period of 180~$^{\circ}$ or 4-fold symmetry with a period of 90~$^{\circ}$. 

(2) With the vertically placed crystal, the polarization axes of 2-fold modes are different than the polarization axes of  4-fold modes.
The 2-fold modes P$_{1}$ and P$_{21}$ are polarized in the same direction, 140~$^{\circ}$. The modes P$_{8}$ and P$_{25}$ are also 2-fold and are polarized in the direction 50~$^{\circ}$. 
The 4-fold modes P$_{2/3}$, P$_{9/10}$ and P$_{22}$ are polarized in the directions 30~$^{\circ}$ and 120~$^{\circ}$. 

(3) The P$_1$, P$_8$, and P$_{25}$ phonon modes are more weakly polarized in the vertically placed crystal than in the horizontally placed crystal.

\begin{figure}[!h]
    \centering
    \includegraphics[width=1\linewidth]{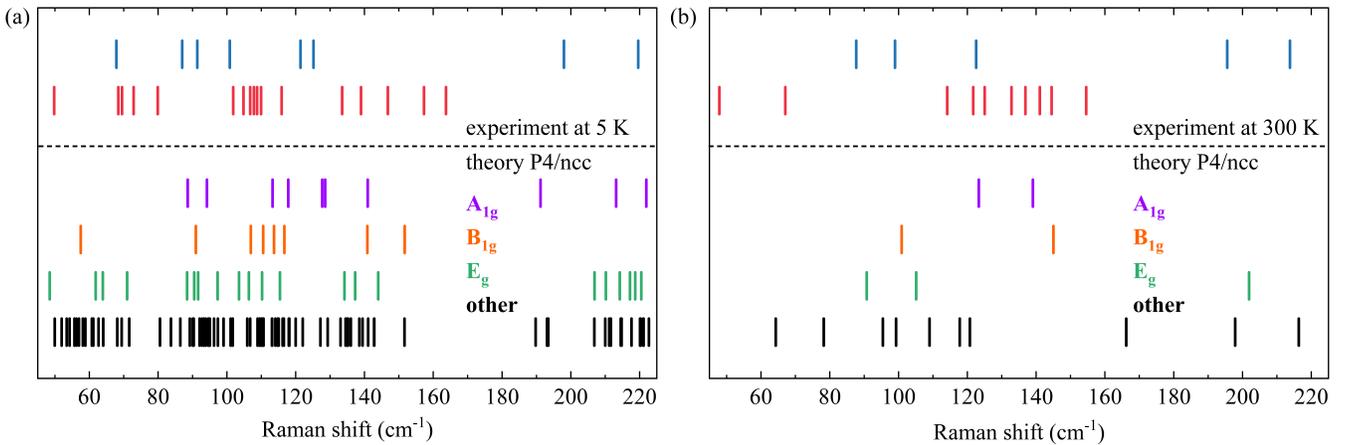}
    \caption
    {\label{SI_experiment_vs_theory.png}
  (a) Comparison between the energy positions of phonon modes observed experimentally at 5~K and those calculated using DFT calculations for the P4/ncc phase (low-temperature structure). (b) Comparison between the energy positions of phonon modes observed experimentally at 300~K and those calculated using DFT calculations for the P4/mcc phase (room-temperature structure). Blue and red lines are energies of modes polarized along and perpendicular to $c$ crystallographic axis, respectively. }
\end{figure}

We attempted to assign the observed phonon modes to their respective symmetries by comparing the experimental Raman energies with those calculated using Density Functional Theory (DFT). This comparison is presented in Fig.~\ref{SI_experiment_vs_theory.png}. We compared the experimental data at 5~K with the DFT results for the P4/ncc phase (panel (a)) and the experimental data at 300~K with the DFT results for the P4/mcc phase (panel (b)). It should be noted that since DFT calculations assume a temperature of 0~K, the theoretical phonon energies are expected to be slightly higher (blueshifted) relative to the experimental values due to the absence of thermal expansion and anharmonic effects. 

The energies of the phonon modes extracted from the experiment are presented in the top parts of the figures, while the energies predicted by theory are presented in the bottom parts. Blue and red lines represent the energies of modes polarized along and perpendicular to the crystallographic $c$ axis, respectively. Purple, orange, and green lines are the energies of A$_{1g}$, B$_{1g}$, and E$_{g}$ modes predicted by theory to be Raman-active in the backscattering geometry. Black lines are the energies of phonon modes with different symmetries or infrared active. 

The RS spectrum at 5~K consists of 25 Raman peaks, while the theoretically predicted phonon modes active in this backscattering configuration should be 64. At room temperature, the situation is reversed: 15 phonon modes are observed, while fewer Raman-active modes (10) are predicted. 
It suggest that we observe non Raman-active modes in our spectrum due to resonant conditions.\cite{Golasa2014} Given the high complexity of the Raman spectrum and the high density of predicted modes, an unambiguous symmetry assignment for each individual mode remains elusive.

\clearpage